\documentclass[useAMS,usenatbib]{mn2e}
\usepackage{amsmath}                       

\usepackage{amsfonts}
\usepackage{graphicx}
\usepackage{float}
\usepackage{morefloats}
\usepackage{listings}
\usepackage{color,hyperref}
\definecolor{linkcolor}{rgb}{0,0,0.25}
\hypersetup{
  colorlinks=true,        % false: boxed links; true: colored links
  linkcolor=linkcolor,    % color of internal links
  citecolor=linkcolor,    % color of links to bibliography
  filecolor=linkcolor,    % color of file links
  urlcolor=linkcolor      % color of external links
}

\newcommand{\kms}{\ensuremath{\mathrm{km}~\mathrm{s}^{-1}}}
\newcommand{\masyr}{\ensuremath{\mathrm{mas}~\mathrm{yr}^{-1}}}
\newcommand{\pc}{\ensuremath{\mathrm{pc}}}
\newcommand{\kpc}{\ensuremath{\mathrm{kpc}}}

\definecolor{kvjcolor}{rgb}{0.75,0,0}

\definecolor{apwcolor}{rgb}{0,0.75,0.75}

\definecolor{jhcolor}{rgb}{0,0,0.75}

\definecolor{edcolor}{rgb}{0,0.75,0.0}

\title[The phase spiral in DR3]
{Multiple phase spirals suggest multiple origins in {\it Gaia} DR3}
\author[J. A. S. Hunt et al.]
  {\parbox{\textwidth}{Jason A. S. Hunt$^{1}$\thanks{E-mail: jhunt@flatironinstitute.org}, Adrian M. Price-Whelan$^1$, Kathryn V. Johnston$^{2,1}$, Elise Darragh-Ford$^{3,4}$}\vspace{0.5cm}
\\
$^{1}$ Center for Computational Astrophysics, Flatiron Institute, 162 5th Av., New York City, NY 10010, USA\\
$^2$ Department of Astronomy, Columbia University, New York, NY 10027, USA\\
$^3$ Kavli Institute for Particle Astrophysics and Cosmology and Department of Physics, Stanford University, Stanford, CA 94305, USA\\
$^4$ 2SLAC National Accelerator Laboratory, Menlo Park, CA 94025, USA
}
\date{Submitted to MNRAS letters, $13^{th}$ June 2022.} 

\pagerange{\pageref{firstpage}--\pageref{lastpage}}
\pubyear{2022}

\begin{document}

\maketitle

\label{firstpage}

\begin{abstract}
{\it Gaia} Data Release 2 (DR2) revealed that the Milky Way contains significant indications of departures from equilibrium in the form of asymmetric features in the phase space density of stars in the Solar neighborhood. 
One such feature is the $z$--$v_z$ phase spiral, interpreted as the response of the disk to the influence of a perturbation perpendicular to the disk plane, which could be external (e.g., a satellite) or internal (e.g., the bar or spiral arms). In this work we use {\it Gaia} DR3 to dissect the phase spiral by dividing the local data set into groups with similar azimuthal actions, $J_\phi$, and conjugate angles, $\theta_\phi$, which selects stars on similar orbits and at similar orbital phases, thus having experienced similar perturbations in the past. These divisions allow us to explore areas of the Galactic disk larger than the surveyed region. The separation improves the clarity of the $z$--$v_z$ phase spiral and exposes changes to its morphology across the different action-angle groups. In particular, we discover a transition to two armed `breathing spirals' in the inner Milky Way. We conclude that the local data contains signatures of not one, but multiple perturbations with the prospect to use their distinct properties to infer the properties of the interactions that caused them.
\end{abstract}


%%%%%%%%%%%%%%%%%%%%%%%%%%%%%%%%  INTRODUCTION   %%%%%%%%%%%%%%%%%%%%%%%%%%%%%%%%%%%
\section{Introduction}
The European Space Agency’s Gaia mission \citep{GaiaMission} has revolutionised our view of the Milky Way. In particular, the second data release \citep[DR2;][]{DR2} highlighted several signatures of dynamical disequilbrium such as ridges in the $R-v_{\phi}$ plane \citep[e.g.][]{KBCCGHS18} and a striking spiral pattern in the $z$--$v_z$ plane, discovered by \cite{Antoja+18}. 

This $z$--$v_z$ `phase spiral' is caused by incomplete vertical phase mixing, following some perturbation to the Milky Way's disc. It has been proposed that an interaction with a perturbing satellite, such as the Sagittarius dwarf galaxy (Sgr) could be responsible \citep[e.g.][]{Antoja+18,Darling+19}, and several simulations of dwarf-Milky Way mergers can qualitatively reproduce the phase spiral \citep[e.g.][]{LMJG19,Khanna+19,Gandhi+20,BlandHawthorn2021,Hunt+21}. 

Attempts to quantitatively reproduce the phase spiral, and the local vertical asymmetry with Sgr alone have been unsuccessful \citep[e.g.][]{BBH21}, and the low present day mass inferred from observations of its remnant \citep{Vasiliev+20} is at odds with the degree of observed disequilibria. However, \cite{BlandHawthorn2021} show that an earlier passage of Sgr when it was more massive can create long lasting phase spirals and disequilibria. Alternatively, \cite{Garcia-Conde+22} find similar levels of disequilibria in a cosmological simulation with several lighter satellites, showing that a more complex dynamical history or the presence of gas dynamics may also help.

\begin{figure}
\centering
\includegraphics[width=\hsize]{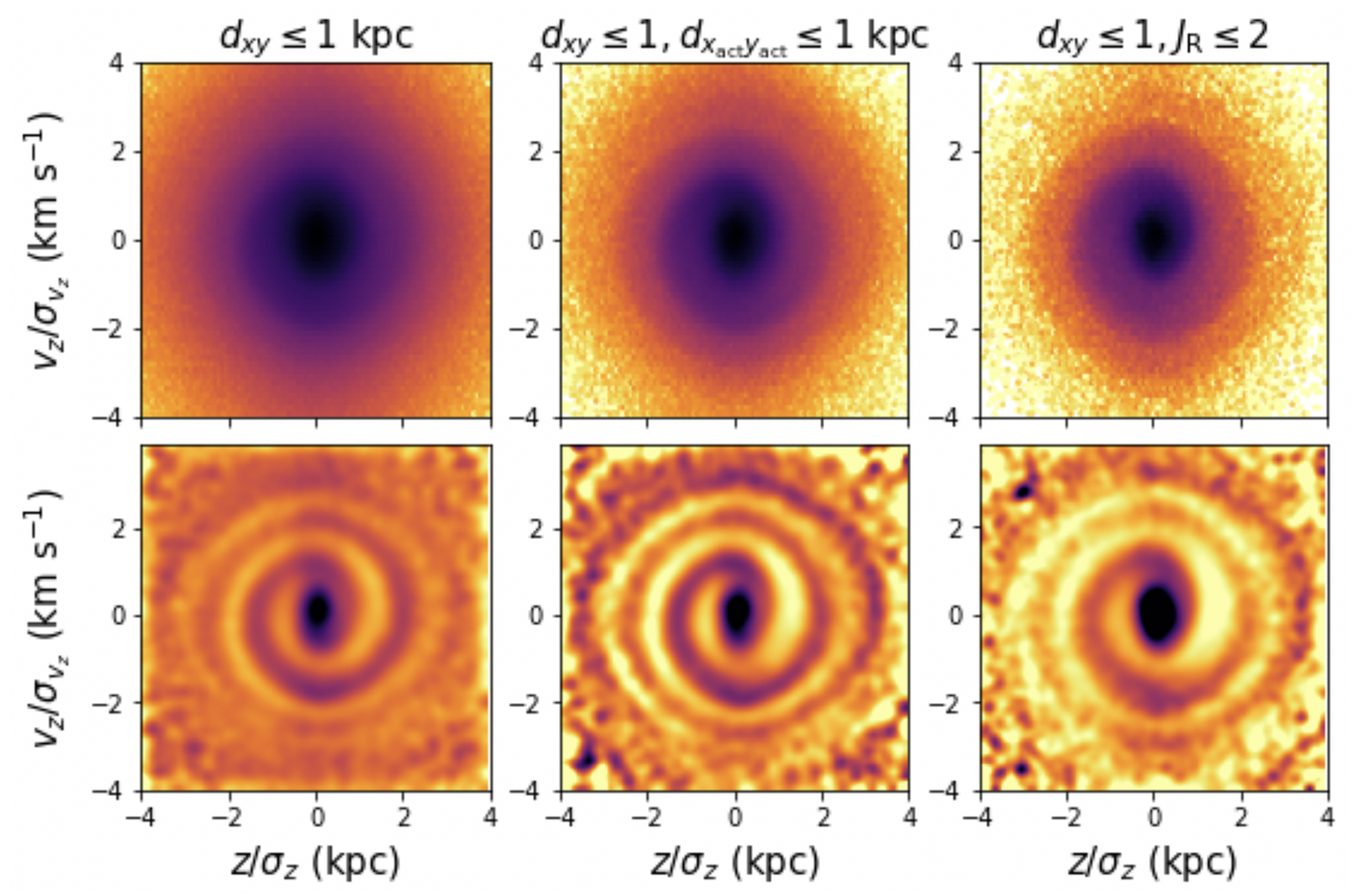}
\caption{The \textit{Gaia} $z$--$v_z$ phase spiral in number counts (upper row) and a Gaussian smooth and subtract representation (lower row) for; \textbf{Left:} all stars within $d_{xy}\leq1$ kpc from the Sun. \textbf{Middle:} stars within $d_{xy}\leq1$ kpc from the Sun, which are also dynamically local (see text). \textbf{Right:} stars within $d_{xy}\leq1$ kpc, with radial action $J_{\mathrm{R}}\leq2$ kpc km s$^{-1}$.}
\label{local}
\end{figure}

Regardless of the origin, the $z$--$v_z$ phase spiral contains information about the nature of the perturbation, and the potential of the Milky Way \citep[e.g.][]{WidmarkII}. In the Milky Way there is a limit to how far we can resolve the phase spiral, yet simulations show clearly that the phase spiral should differ across the Galaxy \citep[e.g.][]{LMJG19,Bland-Hawthorn+19,Hunt+21,Gandhi+20}, both as a consequence of the stellar vertical frequencies across the disc, and also how different stars felt the original perturbation. This implies that by exploiting phase spirals across the Galaxy, we can map the potential spatially \citep[e.g.][]{WidmarkIV}, and also plausibly recover the details of the perturbation. 

\begin{figure*}
\centering
\includegraphics[width=\hsize]{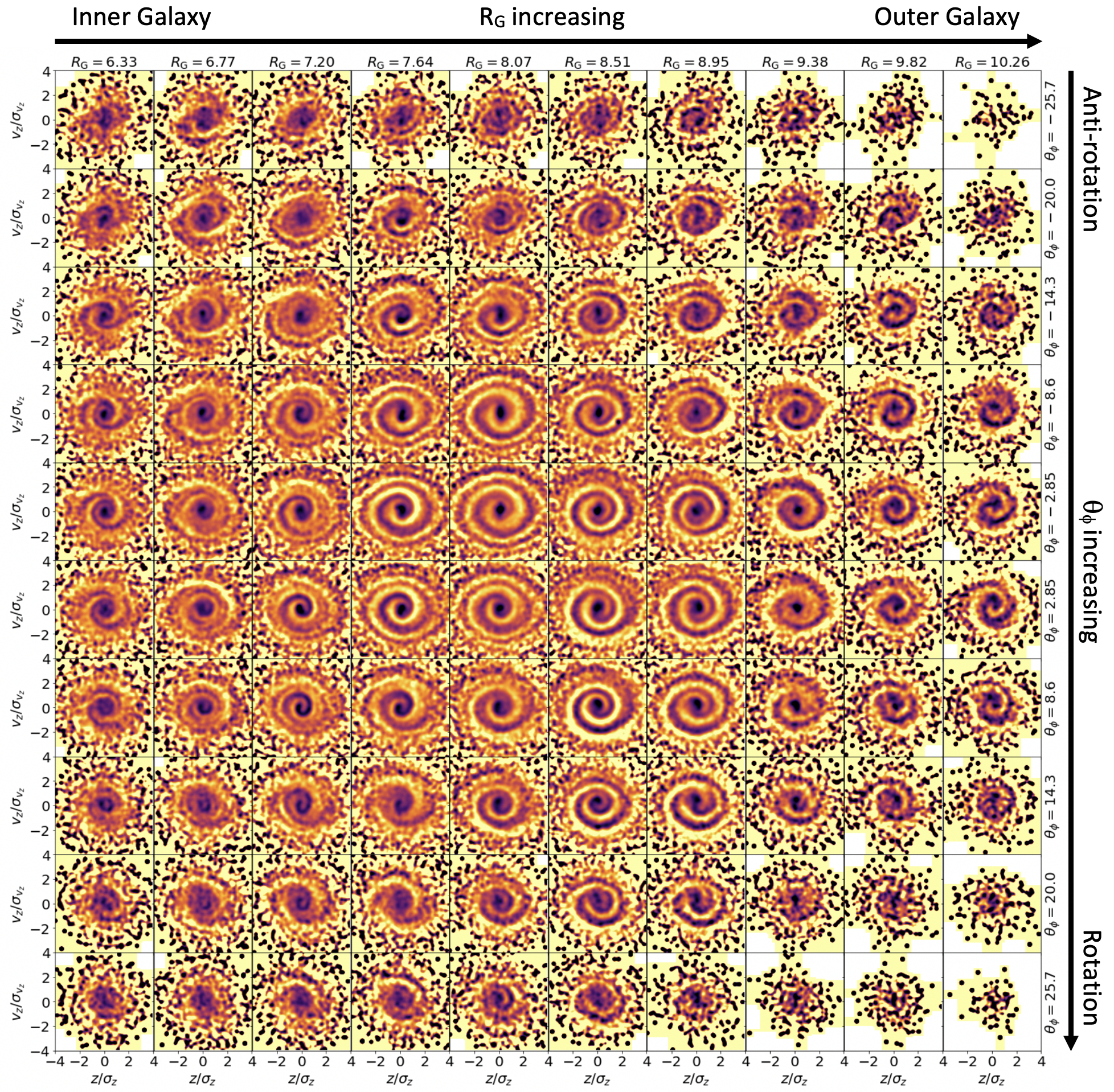}
\caption{$z$--$v_z$ planes for a sample of stars within $d_{xy}<1$ kpc the Sun, split by $R_{\mathrm{G}}$ (kpc, different columns) and $\theta_{\phi}$ (deg, different rows) into dynamically local samples. Note the smooth change in morphology across nearby panels, and the transition from one armed spirals in the majority of the figure, to two armed spirals in the left two columns.}
\label{SplitSpirals}
\end{figure*}

\textit{Gaia} DR3 \citep{DR3} increases the completeness of the sample of stars with 6D phase space information, extending the region over which we can trace the spiral \citep[see][]{DR3-chemistry}. However, the largest sample of highest quality data with complete phase-space information is still dominated by a volume local to the Sun. By splitting this local volume in azimuthal action, $J_{\phi}$, we can explore orbit space over a range of guiding radii \citep[e.g.][]{Li20,Gandhi+20} and by splitting by the conjugate azimuthal phase angle, $\theta_{\phi}$ \citep{Hunt+21}, we can explore the orbit space ahead and behind the Solar azimuth. These divisions effectively place stars into groups --- hereafter referred to as `dynamically local' --- that approximately travel together (i.e. have similar orbital time periods and are currently at the same phase) around the disk which will experience perturbations in the same way. On the other hand, such projections induce extra selection effects, which must be taken into account \citep{Hunt+20}.

In this letter, we show the $z$--$v_z$ phase spiral with the increased coverage of the \textit{Gaia} DR3 radial velocity sample, and illustrate the benefits and limitations of splitting the spiral by actions and angles. In particular, we illustrate the change in phase spiral across the disc, including the first discovery of a two armed `breathing spiral' in the Milky Way (\ref{twoarms}), and present a simple model to reproduce it (\ref{Model}).

%%%%%%%%%%%%%%%%%%%%%%%%%%%%%%%%  The Data  %%%%%%%%%%%%%%%%%%%%%%%%%%%%%%%%%%%
\section{The DR3 data}\label{data}
In this section we describe our treatment of the \textit{Gaia} DR3 data, and the transformation to actions and angles. For our parent sample, we select all stars from the \textit{Gaia} DR3 catalogue that have radial velocity information and a relative parallax error ($\sigma_\varpi$) less than 20 per cent (i.e. $\varpi / \sigma_\varpi > 5$). For this sample, we transform from the \textit{Gaia}-observed (solar system barycentric) ICRS reference frame to Galactocentric Cartesian and cylindrical coordinates \citep{astropy:2018} using a sun--Galactic centre distance of $R_0=8.275~\kpc$ \citep{GRAVITY:2021}, a Solar height above the midplane of $20.8~\pc$ \citep{BB19}, a Sgr A$^*$ radial velocity of $-8.4~\kms$ \citep[this value is computed using the reported $z_0^\cdot=-2.6~\kms$ and adding their fiducial value of 11~\kms;][]{GRAVITY:2021}, and a proper motion of Sgr A$^*$ (in the ICRS frame) of $\boldsymbol{\mu_{\rm ICRS}} = (-3.16, -5.59)~\masyr$ \citep{Reid:2020}. Together, these values imply a total Solar velocity with respect to the Galactic center of $\boldsymbol{v}_\odot = (8.4, 251.8, 8.4)~\kms$.

To compute actions, $\boldsymbol{J} = (J_R, J_\phi, J_z)$, and their conjugate phase angles, $\boldsymbol{\theta} = (\theta_R, \theta_\phi, \theta_z)$, we adopt a  four-component Milky Way mass model consisting of spherical Hernquist nucleus and bulge components, an (approximate) exponential disk component, and a spherical NFW halo component.
We use an approximation of an exponential disk density model \citep{Smith:2015} to represent the Milky Way disk, adopting a radial scale length $h_R=2.6~\kpc$ and scale height $h_z=300~\pc$ \citep{Bland-Hawthorn:2016}.
We fit for the masses and scale radii of the nucleus, bulge, and halo components using the same compilation of Milky Way enclosed mass measurements as used to define the \texttt{MilkyWayPotential} in \texttt{gala} \citep{gala}, with the additional constraint of having a circular velocity at the solar position $v_c(R_0) = 229~\kms$ \citep{Eilers:2019}.
We then compute actions using the ``St\"ackel Fudge'' \citep{Binney:2012, Sanders:2012} as implemented in \texttt{galpy} \citep{galpy}.

%%%%%%%%%%%%%%%%%%%%%%%%%%%%%%%%  The Spirals  %%%%%%%%%%%%%%%%%%%%%%%%%%%%%%%%%%%
\section{The new view of the $z$--$v_z$ phase spirals}\label{spirals}
In this section we explore how the $z$--$v_z$ phase spiral changes with orbital properties by making selections on action and angles from our parent sample of \textit{Gaia} stars. As an initial illustration, the left column of Figure \ref{local} shows the $z$--$v_z$ phase spiral for all `physically local' stars within a cylindrical distance, $d_{xy}\leq1$ kpc of the Sun, for comparison with later plots. The axis ratio is chosen such that the spiral is approximately circular, by dividing by the standard deviation in each coordinate. The top row shows the logarithmic number counts, and the lower row shows a Gaussian-smoothed, background-subtracted representation as shown in past work \citep[e.g.][]{LMJG19,Hunt+21}. 

Regardless of the projection, there is an increase in definition compared to the view from \textit{Gaia} DR2 or eDR3 \citep[e.g.][]{Antoja+18}. The middle column of Figure \ref{local} shows a further selection to stars which are also `dynamically local', which we define as being within $d_{x_{\mathrm{act}}y_{\mathrm{act}}}\leq1$ kpc of the Sun in the action-angle Cartesian projection, $X_{\mathrm{act}}=R_{\mathrm{G}}\cos{\theta_{\phi}}$, $Y_{\mathrm{act}}=R_{\mathrm{G}}\sin{\theta_{\phi}}$ \citep[as shown in][]{Hunt+20}. Note the increased clarity as this is now a sample of stars on similar orbits, which should respond similarly to a perturbation, instead of stars which happen to be physically coincident. However, this introduces an implicit selection on other actions and phase angles, as discussed below. Similarly, the right column of Figure \ref{local} shows the phase spiral for stars within $d_{xy}\leq1$ kpc, now with an explicit radial action selection of $J_{\mathrm{R}}\leq2$ kpc km s$^{-1}$, which results in a similar spiral to that in the middle column, as it should, because stars with low radial action have their guiding centres close to their physical location. 

Along with this refined view of the local $z$--$v_z$ phase spiral, we can also perform the projection of a local physical selection into phase spirals separated in orbit space, as shown previously in \cite{Hunt+21} for a simulation of a merger between a dwarf galaxy and a disc galaxy. In Figure \ref{SplitSpirals} we show stars that are physically located within $d_{xy}\leq1$ kpc of the Sun, split by guiding radius, $R_{\mathrm{G}}$ (from left to right moves from the inner to outer galaxy) and by azimuthal phase angle, $\theta_{\phi}$ (from top to bottom goes in the direction of Galactic rotation) into many `dynamically local' groups. 

This is not entirely equivalent to looking at the $z$--$v_z$ phase spirals of stars that are physically located in other parts of the galaxy, but it does allow us to examine the phase spirals of stellar populations that spend most of their time away from the Solar neighborhood, whether in the inner, or outer Galaxy, without having to deal with higher observational errors or extinction. 

This projection illustrates several points. First that we can access information about regions of the Galactic Disk more than four times larger in surface area than is actually explored by the data set, and without the influence of extinction. Second, that the local structure as shown in Figure \ref{local} is actually a superposition of many spirals, with subtly different morphology or amplitude, that changes smoothly as a function of $R_{\mathrm{G}}$ and $\theta_{\phi}$ (see the companion paper; Darragh-Ford et al., in prep., for a thorough illustration).

However, as mentioned above, splitting a local sample by $J_{\phi}$ (or $R_{\mathrm{G}})$ and/or $\theta_{\phi}$ introduces a dependence on other invariants. As we move further from the centre of Figure \ref{SplitSpirals}, stars that lie outside 1 kpc from the center cannot have zero radial action and still appear within the physically local volume of $d_{xy}\leq1$ kpc, and the further from the central panels, the larger $J_{\mathrm{R}}$ a star must have to be in the sample. 

Thus, Figure \ref{SplitSpirals} is only showing the phase spirals for stars on eccentric orbits in the panels further from the `dynamically local' Solar neighbourhood. Conversely, highly eccentric stars that should have their guiding centre in the central panels may be absent, because their current orbital phase takes them beyond our observable sphere, and thus stars must have sufficiently low $J_{\mathrm{R}}$ to appear in the central panels. Stars in the outer panels will also be collections of stars with similar $\theta_{\mathrm{R}}$, because they only appear in the physically-local volume at certain radial epicyclic phases. 

Most strikingly, Figure \ref{SplitSpirals} shows evidence of a radical transition in spiral morphology. Specifically, the left columns of Figure \ref{SplitSpirals}, with $R_{\mathrm{G}}\approx6.33$ contains the first discovery, to our knowledge, of two armed `breathing spirals' in the Milky Way, as previously only shown in simulations \citep{Hunt+21,BBH21}, and theoretically \citep[][Banik et al., submitted]{Candlish+14}. This detection of two distinct families of phase spiral is exciting for future attempts to recover the history of perturbations to the Galactic disc.

\subsection{The two armed `breathing' spirals}\label{twoarms}
In \cite{Hunt+21} we showed that a breathing mode in a galactic disc \citep[e.g.][]{Widrow+14} manifests as a two armed $z$--$v_z$ phase spiral. In this previous model we found such features primarily in the outer disc in contrast with the \textit{Gaia} data. The presence of two armed spirals in the inner Galaxy suggests that the inner Milky Way contains a breathing mode, as noted previously by \cite{Williams+2013}, and \cite{Carrillo+18} using data from RAVE \citep{Sea06} and \textit{Gaia} DR2.

Conversely, the one armed spirals in the Solar neighborhood, and a few kpc outwards arise from a bending mode. Such a combination can be physically intuitive, if the breathing mode in the inner Galaxy and the bending mode in the outer galaxy have two different origins. For example, it has been shown that a breathing mode can be induced in the inner Galaxy by both bars and spiral arms \citep[e.g.][]{Monari+2015,Monari+2016a}. Thus, this dramatic shift in morphology could be telling us that we are looking at the transition between $z$--$v_z$ phase spirals that have originated internally \citep[e.g.][]{Khoperskov+19-buclking}, and from some external perturbation \citep[e.g.][]{Antoja+18,Laporte+18a,Hunt+21}. However, it has also been noted that the strength of the breathing mode in the inner Galaxy is higher than would be expected from purely internal excitation \citep{Carrillo+19}. Other options include a longer lived perturbation of the Galactic disc by structure in the dark matter halo, and/or relics of earlier mergers in the Milky Way's history, which is a topic of future work.

\subsection{Recreating the two armed `breathing' spirals}
\label{Model}
In order to reproduce such breathing spirals, we set up an $N$-body model, which is a reproduction of Model A from \cite{Hunt+20} (see within for parameters), re-simulated with a halo with $8\times10^8$ dark matter particles, a thin disc with $2\times10^8$ particles, and two thick discs with $3\times10^7$ and $1\times10^7$ particles each. The galaxy forms a bar and spiral structure in isolation. We then perform a second model where this host also undergoes an interaction with a satellite. Both are evolved for 3 Gyr using \texttt{Bonsai} \citep{Bonsai}.

\begin{figure}
\centering
\includegraphics[width=\hsize]{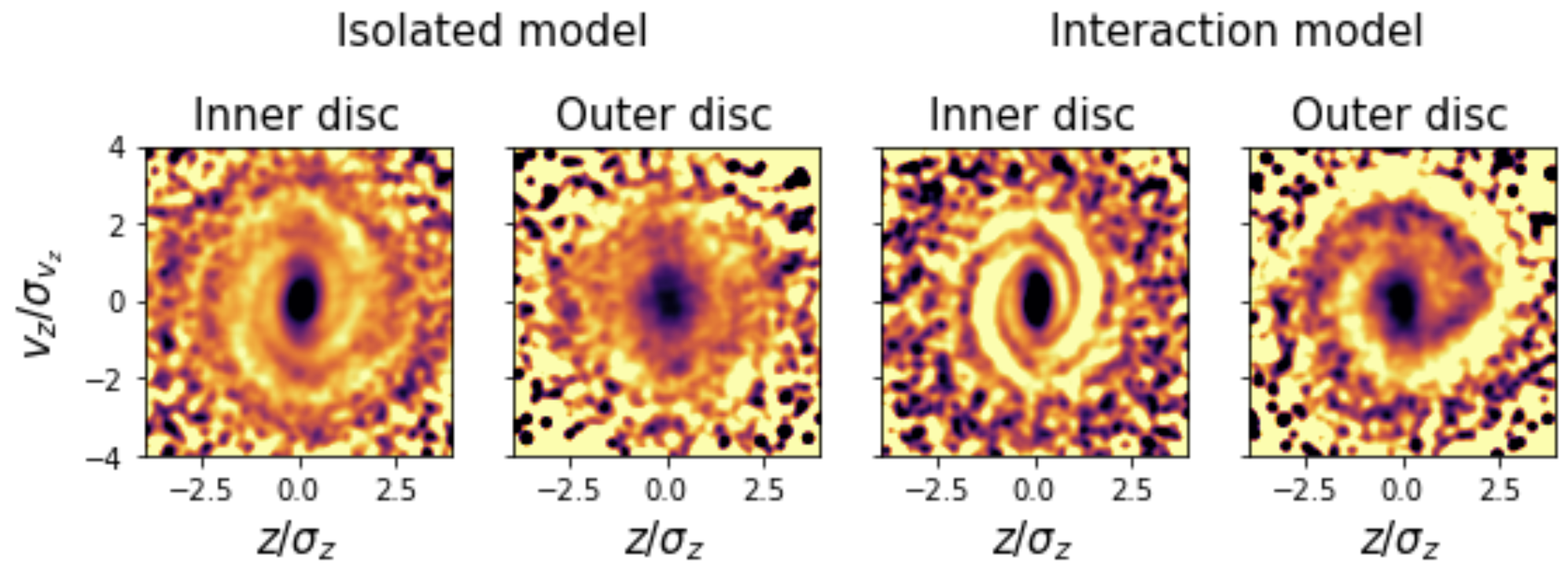}
\caption{\textbf{Left:} Two armed spiral from the inner disc of the isolated galaxy model. \textbf{Middle left:} No spiral in the outer disc of the isolated galaxy. \textbf{Middle Right:} Two armed spiral remains present in the galaxy which interacts with a satellite. \textbf{Right:} One armed spiral in outer galaxy induced by the interaction.}
\label{ModelPlot}
\end{figure}

The isolated galaxy forms a bar and spiral structure through secular evolution. Interestingly, this isolated case forms a series of two armed `breathing spirals' across the disc, that appear, disappear and reappear across the face of the disc as the galactic structure forms and evolves. The left hand panel of Figure \ref{ModelPlot} shows an example of a breathing spiral from the inner part of the isolated galaxy ($R_{\mathrm{G}}=6$ kpc) at $t=1.66$ Gyr. Note that this formation of $z$--$v_z$ phase spirals in an isolated disc is distinct from the bar buckling origin presented in \cite{Khoperskov+19-buclking}, as while our model contains a bar, it does not buckle, and we see no one armed `bending spirals' over the 3 Gyr evolution. The second panel of Figure \ref{ModelPlot} shows that there is no phase spiral in the outer disc ($R_{\mathrm{G}}=12$ kpc) of the isolated galaxy.

For the externally-perturbed galaxy, the satellite passes through the disc at $t=1.14$ Gyr, which excited one armed $z$--$v_z$ phase spirals in the outer galaxy, as seen in previous works. The third panel of Figure \ref{ModelPlot} shows a similar two armed phase spiral at the same position and time as seen in the isolated case in the left panel. The right hand panel of Figure \ref{ModelPlot} shows a one armed phase spiral in the same snapshot, located further out in the galaxy, which was not seen in the same position in the isolated galaxy (second panel). This suggests that the satellite directly excites a bending mode in the outer disk. However, the self-sustained breathing mode spirals in the inner disk are present regardless.

While we have made no attempt to quantitatively match the \textit{Gaia} data, or the morphology of the interaction, the resulting $z$--$v_z$ phase spirals are similar to the \textit{Gaia} data, with the transition between a $m=1$ and $m=2$ spiral appearing inside the Solar radius, suggesting that we may be seeing a transition between internally- and externally-excited phase spirals just inside the Solar neighborhood.

%%%%%%%%%%%%%%%%%%%%%%%%%%%%%% SUMMARY %%%%%%%%%%%%%%%%%%%%%%%%%%%%%%%%%%
\section{Summary}
\label{Summary}
This paper presents a new view of the $z$--$v_z$ phase spirals in \textit{Gaia} DR3 by dividing the a local data set by azimuthal action and angle. We show the extent of orbit space that the Solar neighborhood sample can span and demonstrate the biases in radial action and angle that this grouping produces.

We report the first detection of the transition from one armed `bending mode' spirals in the Solar neighbourhood, to two armed `breathing spirals' in the inner Milky Way, supported by prior measurements of the transition between a breathing mode and a bending mode in the disc, just inside the the Solar circle. While previous modelling of a satellite impact on a Milky Way like disc galaxy has shown two armed `breathing spirals' in the outer galaxy \citep{Hunt+21}, or across a wide variety of radii \citep{BBH21}, it is expected that they should arise from any vertical breathing mode (Banik et al., submitted), regardless of the origin.

\nocite{BanikSpirals1}

We recreate such two armed `breathing spirals' in a $N$-body simulation of the Milky Way through secular evolution of a galactic bar and spiral structure, which is distinct from the bar bucking scenario of \cite{Khoperskov+19-buclking}, because our bar does not buckle. Perturbing this disc with a satellite then creates a split of two armed `breathing spirals' interior to the Solar radius, and one armed `bending spirals' outside the solar radius, as seen in the data. 

Further modelling is required to better fit this transition, and to map the specific features of the phase spirals and quantitatively link them to a specific origin. We defer this to the companion paper; Darragh-Ford et al. (in prep.), and future work. For now, our limited analysis demonstrates the depth of information on the history of perturbations to the Galactic disk contained in local data sets.

\nocite{Elise1}

\vspace{-4pt}

\section*{Acknowledgements}
We thank the anonymous referee for a constructive report. We make use of data from the European Space Agency (ESA) mission \textit{Gaia} (\url{http://www.cosmos.esa.int/gaia}), processed
by the \textit{Gaia} Data Processing and Analysis Consortium (DPAC,
\url{http://www.cosmos.esa.int/web/gaia/dpac/consortium}). Funding for the DPAC has been provided by national institutions, in particular the institutions participating in the \textit{Gaia} Multilateral Agreement.
KVJ's contributions were supported by NSF grant AST-1715582. This project was developed in part at the Gaia F\^ete, held at the Flatiron institute Center for Computational Astrophysics in 2022 June.
\vspace{-4pt}

\section*{Data availability}
The $Gaia$ data is available at {\url{https://gea.esac.esa.int/archive/}}
\vspace{-4pt}
\bibliographystyle{mn2e}
\bibliography{ref2}

\label{lastpage}
\end{document}